\documentclass[a4paper,11pt]{article}
\usepackage{pos}

\title{How to predict a new physics scale and its uncertainty with the Inverse Amplitude Method}

\author[a]{Alexandre Salas-Bern\'ardez\footnote{Speaker. The author thanks his collaborators on this project F. J. Llanes-Estrada, J. Oller and J. Escudero-Pedrosa. This work was funded by MINECO:FPA2016-75654-C2-1-P, MICINN: PID2019-108655GB-I00, PID2019-106080GB- C21, UCM: EB15/21 (Spain).}}

\affiliation[a]{Universidad Complutense de Madrid, Dept. de F\'isica Te\'orica and IPARCOS,\\
28040 Madrid, Spain. (On absence at University of Graz)}

\emailAdd{alexsala@ucm.es}

\abstract{Effective Field Theories such as HEFT, organized as momentum expansions, are a controllable approximation to strong dynamics only near threshold, as they miss exact elastic unitarity, reducing their predictive power at a higher scale if small separations from the Standard Model are found at the LHC or elsewhere. Unitarized chiral perturbation theory extends their reach to the saturation of unitarity but, generally with unknown systematics. Here I present a brief 
"handbook" on how to put to use the Inverse Amplitude Method to predict a resonance (associated with a new physics scale in Beyond the Standard Model searches) and more especially on how to control its systematic uncertainties.}

\FullConference{%
  *** The European Physical Society Conference on High Energy Physics (EPS-HEP2021), ***\\
  *** 26-30 July 2021 ***\\
  *** Online conference, jointly organized by Universität Hamburg and the research center DESY ***
}

\begin{document}
\maketitle

\section{Introduction}
It is seeming more and more likely that the LHC or other current accelerators will not find direct evidence of new particles. This is why a huge effort on studying the traces that the new physics leaves on the visible particle content is being made. This study relies naturally on Effective Field Theories (EFTs). However, the naive or direct use of an EFT will likely be of no use for predicting the behaviour of scattered particles above the energies hitherto reached by accelerators. This is so because an EFT scattering amplitude does not respect unitarity (an illustration of this fact is shown in Fig. 1). Several unitarization methods have been devised to solve this problem, such as the K-matrix method; the $N/D$ method; variations of the Bethe-Salpeter equation or the Inverse Amplitude Method (IAM) \cite{Truong:1988zp}. The difference between them is purely empirical, since, usually, their predictions can only be tested if higher energies are reached. This is why it appears useful to control the systematic uncertainties of each method \textit{ab initio}, so that their predictions are also given within an error-band.

\begin{figure}[ht!]
\centering
\includegraphics[width=90mm]{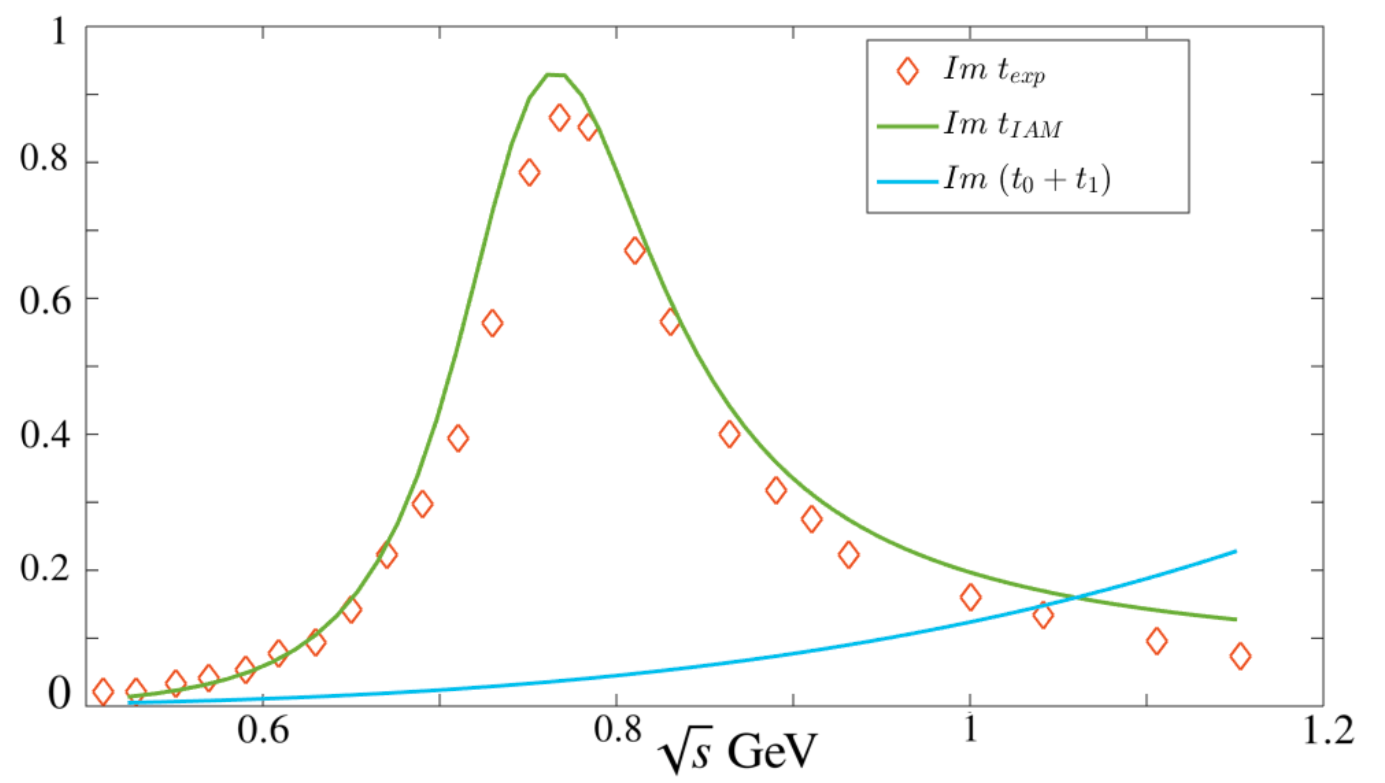}
\caption{Prediction of the $\rho$ resonance with the IAM using only threshold $\pi\pi$-scattering data. The pure EFT amplitude (in blue) completely misses the resonance and fails to reproduce the experimental data (in orange, taken from \cite{GomezNicola:2001as}) just above threshold. The IAM predicts the position of the resonance with an error of about 10\%.}
\end{figure}

All things said, the most successful unitarization methods in the market also respect the analyticity properties of amplitudes, this ensures causality in the effective theory (the “old” K- Matrix method does not). Among these methods, the IAM is salient due to its simple and algebraic form (the $N/D$ method is based on integral equations and large-$N_C$ methods are only approximate for certain symmetry breaking patterns). This is why, in \cite{Salas-Bernardez:2020hua}, we have laid out the systematic uncertainties of the IAM based on a dispersive/analytic approach.\\
In this Proceedings, I try to make a clear summary on how to put to use the IAM to obtain a (resonant) new physics scale and its theoretical uncertainty from LHC scattering data (f.e. the scattering of $W_LW_L$ in Higgs EFT). This exercise is relevant to assess if, given precision data is obtained at the LHC or elsewhere, whether the new physics scale may be at the reach of future colliders (like FCC) or not. In Section \ref{IAMuncert} I will present an "IAM Uncertainties Handbook", \textit{i.e.} a quick list of the uncertainties of the method and how to calculate them if wanted.

\section{Inverse Amplitude Method's Uncertainties Handbook}\label{IAMuncert}
Amplitudes of two-to-two-body scattering for identical particles and definite isospin $I$, 
$T_I(s, t, u)$ (expressed in terms of the usual Mandelstam’s variables $s$, $t$ and $u$), admit a low-energy simplification: the partial-wave projection
\begin{equation}
T_I(s,t,u)=32 \pi\sum_{J=0}^\infty(2J+1)t_{IJ}(s)P_{J}(\cos\theta_s)\;.
\end{equation}
The partial waves, $t_{I J} (s)$, are easily tractable in the context of EFTs since they can be Taylor-expanded
at low energies as $t(s) = t_0(s)+t_1(s)+...$\footnote{Here and in the rest of the paper I omit spin and isospin indices.}, here only the terms up to NLO ($\mathcal{O}(p^4) = \mathcal{O}(s^2)$) are made
explicit. Unitarity of the $S$ matrix forces the partial waves to follow the relation $\text{Im } t(s) = \sigma(s)|t(s)|^2$
for physical $\text{Re } (s) > 0$ (here $\sigma(s)=\sqrt{1-\frac{s}{4m^2}}$ is the partial-wave two-body phase space factor for 
particles of mass $m$ and for $s \geq 4m^2$). However the EFT expansion satisfies this relation only order by order, f.e. $\text{Im } t_1(s) = \sigma(s)|t_0(s)|^2$. On the other hand, if one expands the reciprocal or inverse amplitude $1/t = 1/t_0 - t_1/t_0^2 + ...$, the NLO approximation for $t(s)$,

\begin{equation}
t_{\text{IAM}}(s)=\frac{t_0^2(s)}{t_0(s)-t_1(s)}\label{IAM}
\end{equation}
will be found to satisfy unitarity exactly thanks to the simple relation $\text{Im } 1/t(s) = -\sigma(s)$.\\
In \cite{Salas-Bernardez:2020hua}, the uncertainties of the IAM are thoroughly studied. The master formula on which \cite{Salas-Bernardez:2020hua} leans on is based on a dispersive/analytic approach and amounts to
\begin{equation}
G(s)=G(0)+G'(0)s+\frac{1}{2} G''(0) s^2+ PC(G)+\frac{s^3}{\pi}LC[G]+\frac{s^3}{\pi}RC[G],\label{GG}
\end{equation}
where $G(s) \equiv t_0^2(s)/t(s)$ and $PC(G)$ are the pole contributions to the dispersive integral coming from zeroes of $t(s)$\footnote{Here $LC[f](s)=\int_{-\infty}^0  dz\frac{\text{Im } f(z)}{z^3(z-s)}$ and $RC[f](s)=\int_{4m^2}^\infty  dz\frac{\text{Im } f(z)}{z^3(z-s)}$, where $m$ is the mass of the scattered particles.}. With this formula at hand it is possible to derive the IAM amplitude in Eq. (\ref{IAM}) by employing the following approximations on Eq. (\ref{GG}):
\begin{itemize}
\item Taking the NLO approximation of the subtraction constants $G^{(i)}(0) \simeq (t_0 - t_1)^{(i)}(0)$.
\item Neglecting the pole contributions of $G(s)$, i.e. zeroes of $t(s)$. These zeroes are of two types and both appear for real positive $s$: below threshold they are denominated Adler zeroes and above threshold they are called Castillejo-Dalitz-Dyson (CDD) zeroes.
\item Approximating at NLO the imaginary part of $G(s)$ over the Left Cut (LC), $\text{Im }G(s) \simeq -\text{Im }t_1(s)$. (The left cut appears on the whole negative real-$s$ axis due to the partial wave projection, giving a discontinuity in the imaginary part of $t(s)$).
\item Neglecting inelastic channels on the whole Right Cut (RC)\footnote{Examples of these inelastic channels in hadron physics would be $K\bar{K}$ or $2n$-$\pi$ intermediate states for $\pi\pi$ scattering. In HEFT, for $W_LW_L$ scattering the inelastic channels would be $2n$-$W_L$ or $n$-$h$.}. Notice the IAM is exact on the RC if inelastic channels are turned off.
\end{itemize}
The next subsection is devoted to explain how these sources of uncertainty can be estimated.

\subsection{Control of uncertainties}
Each source of uncertainty in the method will affect the prediction for the resonance’s mass and hence the new physics scale predicted by it (see in \cite{Espriu:2021oqp} a comprehensive way of constructing HEFT partial wave amplitudes). The mathematical procedure is to quantify the absolute value of
\begin{equation}
\Delta (s) G(s)\equiv \Big(\frac{t(s)-t_{\text{IAM}(s)}}{t_{\text{IAM}}(s)}\Big)\frac{t_0^2(s)}{t(s)}\;.
\end{equation}
Since it is the above factor that modifies the IAM’s condition for a resonance as
\begin{equation}
t_0(s_R) - t_1(s_R)+i \sigma(s_R)t_0^2(s_R)=\Delta(s_R)G(s_R) \;,
\end{equation}
for real $s_R$ (the resonance’s mass squared). The steps to quantify its magnitude are:
\begin{enumerate}
\item To include the NNLO correction on the subtraction constants, one should use Resonance Effective theory and lean on \cite{Guo:2009hi} (Eq. (A4) there). However these constants are not yet known in HEFT (examples can be found in \cite{Pich:2020xzo}). One should estimate the size of the $\mathcal{O}(s^3)$ terms at the resonance’s squared mass $s_R$ and equate it to $|\Delta(s_R)G(s_R)|$.
\item To assess the uncertainty due to neglecting the Adler zeroes simply compute the ratio $m^4/s^2_R$.
The appearance of CDD zeroes can be checked by solving the simple algebraic equation
$\text{Re }t_1(s_C ) + t_0(s_C ) = 0$ for $s_C$ (zero’s position). If a solution is found, the IAM must fail, but can be easily modified as $t_{\text{IAM}}=\frac{t_0^2}{t_0-t_1}\to \frac{t_0^2}{t_0-t_1+\frac{s}{s-s_c}\text{Re}(t_1)}$. This substitution solves the issue.
\item The easiest (though the less reliable and crude among the ones presented in \cite{Salas-Bernardez:2020hua}) way of assessing the uncertainty on the left cut is simply assigning to its uncertainty the LC integral of $t_1$ since it is presumably an upper bound \cite{Salas-Bernardez:2020hua}, $|\Delta(s_R)G(s_R)| = s^3_R|LC[t_1](s_R)|/\pi$.
\item Inelastic channels fall into two types. The two-body ones can be treated using the matrix form of the IAM (see \cite{GomezNicola:2001as}), and be fully taken into account.
$n$-body intermediate states can be controlled by evaluating the ratio of the $n$-body phase space factor against the two-body one, $R(s) = \phi_n(s)/(v^{2n-4}\phi_2(s))$, at the maximum value where one trusts the EFT unitarization (namely $4\pi v$ for HEFT, where $v = 246$ GeV). Then one equates the displacement of the pole to $|\Delta(s_R)G(s_R)| = R((4\pi v)^2)\cdot s^3_R |RC[t_1](s_R)|/\pi$.
\end{enumerate}
Finally, to compute the displacement of the pole, one has to treat each contribution to $|\Delta(s_R)G(s_R)|$ separately depending on their scaling just as is presented in Appendix A of  \cite{Salas-Bernardez:2020hua}.

\end{document}